\documentclass[prl,nofootinbib,twocolumn,superscriptaddress]{revtex4}

\def\mysection#1{{\bf #1.} }

\def\lsim{\mathrel{\rlap{\lower4pt\hbox{\hskip1pt$\sim$}}
    \raise1pt\hbox{$<$}}}         
\def\gsim{\mathrel{\rlap{\lower4pt\hbox{\hskip1pt$\sim$}}
    \raise1pt\hbox{$>$}}}         

\begin{document}

\begin{titlepage}

\hfill$\vcenter{
\hbox{\bf SU-4252-830} }$ 

\hfill$\vcenter{
\hbox{\bf BNL-HET-06/4} }$

\begin{center}
{\Large \bf  Collider Signals of Top Quark Flavor Violation
from a Warped Extra Dimension}
\vskip .1in
{\bf Kaustubh Agashe}$^*$,  
{\bf Gilad Perez}$^\dagger$
and {\bf Amarjit Soni}$^\#$\\  
{\em $^*$ Department of Physics,
Syracuse University, 
Syracuse, NY 13244}\\ 
 {\em $^\dagger$ Theoretical Physics Group, 
Lawrence Berkeley National Laboratory, Berkeley, CA 94720}\\  
{\em $^\#$ Brookhaven National Laboratory, 
Upton, NY 11973}

\end{center}

\begin{center} {\bf Abstract}\\\end{center}

We study top quark flavor violation 
in the framework of a warped extra dimension with the Standard Model (SM) fields
propagating in the bulk.
Such a scenario provides solutions to both the Planck-weak hierarchy
problem and the flavor puzzle of the SM without inducing a flavor problem.
We find that, generically, $tcZ$  couplings
receive a huge enhancement,
in particular the right handed ones can be ${\cal O} (1 \%)$. 
This results in BR $\left( t \rightarrow c Z \right)$ at or above
the sensitivity of the Large Hadron Collider (LHC). At the 
International Linear Collider (ILC), single top 
production, via $e^+ e^- \to t \bar c$,  
can be a striking signal for this scenario. In particular,
it represents a physics topic of critical importance that can be
explored even with a relatively low energy option, close to the
$tc$ threshold. At both the LHC and the ILC,
angular distributions can probe the above prediction of dominance of
right-handed couplings.

\vskip .2in

\end{titlepage}
\newpage
\renewcommand{\thepage}{\arabic{page}}
\setcounter{page}{1}

\mysection{Introduction}
In a few years, the Large
Hadron Collider (LHC) is expected to unravel the 
mystery of electroweak symmetry breaking (EWSB)
and also perhaps the mechanism of stabilizing the enormous hierarchy
between the Planck and EWSB scales.
Can this TeV-physics give us clues to
the origin of flavors? 
The answer to this question 
depends on the scale of dynamics which mediates flavor physics, $\Lambda_F$. 
It is the top quark contributions to
the Higgs mass squared which yield the most severe fine tuning within
the SM due to large top mass. In almost any natural SM extension,   
therefore, the top quark is likely to
have significant couplings to the new physics (NP) sector at TeV.
Generic couplings of the NP sector to the light quarks are in tension
with the constraints
from flavor changing neutral currents (FCNC) processes which require
the NP scale to be of $O(1000)$ TeV.
However, in
models which have a high $\Lambda_F$, the flavor structure 
in TeV scale physics 
is described entirely by the
up and down Yukawa matrices -- these models belong to the minimal flavor
violation (MFV) framework~\cite{MFV}.
Such a scenario is rather easily consistent with FCNC data
even with TeV NP scale and on the flip side, it is
difficult to obtain clues
to the origin of flavors from the NP at TeV in this case.

However, references ~\cite{nmfv,dmsNMFV} studied a different
possibility that new sources of flavor and CP violation are present
in the NP at TeV.
It was shown that, as long as the NP dynamics respect the SM
approximate flavor symmetries and is
{\em quasi}-aligned (i.e., has at most 
CKM-like misalignment) with SM Yukawa 
matrices,
such a low  flavor scale 
is still allowed by the FCNC data.
The corresponding framework was denoted as next to MFV (NMFV)~\cite{nmfv}.
Thus an exciting case is possible 
in which flavor violation arises from the same NP at TeV scale
which is related to the solution of the hierarchy problem \cite{nmfv}.
All of the precise data constraining this
framework, 
available at present, is due to processes which
involve down type quarks. However,
the most direct way to test the above paradigm is via a careful study
of the top couplings.
For the first time such a test will be possible 
at the LHC since millions of top quarks will be produced per year.
In particular we will mainly focus here on $\Delta F=1$ top 
FCNC processes related to
$ t \rightarrow c $ transition which are highly GIM and CKM-suppressed
within the SM, but yet are theoretically clean due to the fact that the top
decays before being hadronized.

In this letter, we study one such scenario which combines
solutions to
the Planck-weak hierarchy and flavor puzzle, namely the 
Randall-Sundrum (RS1) framework of
warped extra dimension \cite{rs1}.
We show that sizable $tcZ$ coupling
is induced which can lead to observable effects at both
the upcoming LHC and at the proposed 
International Linear Collider (ILC).

The framework 
involves a slice of AdS$_5$. Due to the warped geometry,
the relationship between the $5D$ mass scales
(taken to be of order the $4D$ Planck scale) and those in an
effective $4D$ description depends on the location in the extra dimension.
The $4D$ (or zero-mode) graviton is localized near the ``UV/Planck''
brane which has a Planckian fundamental scale, whereas  
the Higgs sector is localized near the ``IR/TeV'' brane where it
is protected by a warped-down fundamental scale of $\sim$ TeV. This
large 
hierarchy of scales can be generated via a modest-size radius of
the extra dimension. Furthermore, based on the AdS/CFT correspondence
\cite{Maldacena:1997re}, RS1 
is conjectured to be dual to $4D$ composite Higgs models \cite{Arkani-Hamed:2000ds}.

In the RS1 model, the entire SM (including the fermions
and gauge bosons) are assumed to be localized on the TeV brane.
Thus, it provides 
no understanding of the flavor puzzle. Moreover, the higher-dimensional
operators in the $5D$ effective field theory
(from cut-off physics) are suppressed
only by the warped-down scale $\sim$ TeV, giving too large
contributions to FCNC processes and observables related to SM electroweak precision tests (EWPT).

An attractive solution
to this problem is to allow
the SM fields to propagate in the extra dimension \cite{bulkgauge, gn, gp}.
In such a scenario, the 
SM particles are identified with the zero-modes of the $5D$ fields
and the
profile of a SM fermion in the extra dimension
depends on its $5D$ mass parameter.
We can 
then choose to localize 1st and 2nd generation fermions near the Planck brane
so that the 
FCNC's from higher-dimensional operators are suppressed
by scales $\gg$ TeV which is the cut-off
at the location of these fermions~\cite{gp, hs}.
As a bonus, we obtain a solution to the flavor puzzle 
in the sense that hierarchies in the SM Yukawa couplings arise without
introducing hierarchies
in the fundamental $5D$ theory~\cite{gn, gp, hs}:
the 1st/2nd generation fermions
have small Yukawa couplings to Higgs which is localized near the
TeV brane. 
Similarly, 
the top quark can be localized near the TeV brane
to account for its large Yukawa.

In this scenario, there is a 
new source of FCNC's from the couplings of SM fermions
to gauge KK modes since these couplings are non-universal due to the different
profiles for the SM fermions.
However, the gauge KK modes are localized near the TeV brane 
while the light fermions are near the Planck brane and hence 
it can be shown that the non-universal part of these couplings 
are proportional to the SM Yukawa couplings~\cite{gp, hs}.
Thus, most of the couplings to the NP degrees of freedom are small and hierarchical,
leading to the same symmetry structure which
suppresses the SM flavor-violating contributions~\cite{aps}.  
This is in sharp contrast to 
similar models in a flat extra dimension which are problematic since they
require the KK scale $\gsim 1000$ TeV to satisfy
FCNC constraints.   
Since the top Yukawa is large,
we expect FCNC's involving top (and also its partner, $b_L$)
to be sizable, especially
given that the KK scale must be a few TeV based on naturalness.
The gauge KK modes
also give contributions to EWPT:
the constraints from the $S$ and $T$ parameters can be satisfied with 
KK mass scale as low as $\sim 3$ TeV 
if a custodial isospin symmetry is incorporated~\cite{custodial}.

Let us examine the top/bottom sector in detail.
It is clear that both $t_{L,R}$ being near the Planck brane gives too small 
top Yukawa.
On the other hand, $(t,b)_L$ being close to the TeV brane leads to
its coupling 
to KK $Z$
being large and, in turn, results in a 
non-universal shift in its coupling to the SM $Z$ via mixing
of KK $Z$ with zero-mode $Z$~\cite{custodial}:
$
\delta g_Z^{ b_L }  \sim  g_{Z^{\rm KK }}^{ b_L } 
\xi
\frac{ m_Z^2 }{ m_{\rm KK }^2 }$
where $\xi\equiv\sqrt{ \log \left( M_{ Pl } / \hbox{ TeV } \right) }$ 
and $g_{Z^{\rm KK }}^{ b_L }$ is the corresponding non-universal KK $Z$
coupling.
There is also a contribution 
from the exchange of KK modes of the
extra $U(1)$ arising from the extended $5D$ gauge symmetry;
here and below ``KK $Z$'' will represent both these effects.
Such corrections to $Z \bar{b}_L b_L$ coupling 
can be suppressed by 
%
%
suitable
choice of representation
%
%
of top and bottom quarks under the custodial isospin symmetry 
\cite{Zbb}, but in
this paper
we will consider models with the assignment of \cite{custodial}. 
%
%
The constraint from data is that
$\delta g_Z^{ b_L } / g_Z \lsim 1/4 \%$. 
Thus, for few TeV KK scale,
there is a tension between obtaining large top
mass and EWPT (i.e., $Z \bar b_L b_L$ coupling) which can be
relaxed by the following setup:
(i) $(t,b)_L$ quasi-localized near TeV brane so that the shift 
in coupling of $b_L$ to $Z$ is on the edge,
(ii) $t_R$ localized very close to TeV brane
to obtain large top quark mass and (iii) largest dimensionless $5D$
Yukawa, $\lambda_{ 5D }\sim 4$,
consistent with perturbativity.
Note that the resulting coupling of $b_L$ to gauge KK modes (including gluon) 
is comparable to the SM couplings and thus is still
larger than what is expected on the basis of $m_b$ alone
(since it is dictated by the large top mass instead). Thus, we obtain
sizable flavor violation involving $b_L$
which has been studied in~\cite{aps,others1, others2}
along with flavor violation in lepton and light quark
sectors.

In the rest of this paper, we focus 
on {\em top} quark flavor violation since 
as mentioned above, it is likely to be sizeable and in a few years,
the LHC will provide us a copious source of tops.

There is a non-universal shift in the coupling of $t_{ R } $ to $Z$ as above,
except that, due to its profile, the coupling of $t_R$ to gauge KK modes
is enhanced (just like those for the Higgs):
$
g^{ t_R }_{ Z^{ \rm KK } } 
\sim  g_Z \xi \,.$
There is also
a similar size effect from mixing 
(via
the Higgs vev) of zero-mode $t_R$ with KK $t_L$ which then couples
to the $Z$~\cite{snowmass}:
$\delta g_Z^{ t_R } |_{ t_L^{ \rm KK } } \sim \left(  
\lambda_{ 5D } \; v / \sqrt{2} \right)^2 / 
m_{ KK }^2.$
The shift in coupling of $Z$ to $t_L$ is the same as that for 
$b_L$, i.e., smaller.

There are also $4$-fermion operators
generated by the direct exchange of 
KK $Z$, $\gamma$.
We can use the fact
that the coupling of light fermions (for example,
the electron) to these KK modes is suppressed
compared to the SM gauge couplings by $\xi$
to obtain the coefficients of these 
operators. The coupling of the extra $U(1)$ 
gauge bosons to
light fermions
is Yukawa suppressed and hence their exchange  
is negligible.

\mysection{Flavor violation}
The couplings discussed above are in the interaction basis.
Flavor violation arises when we rotate to the mass basis.
To determine these effects, we need to estimate the corresponding mixing angles.

We assume that the $5D$ Yukawas are anarchic 
so that the hierarchies in both the SM fermion masses and mixing angles
orginate from the profiles.
Since $u_L$ and $d_L$ have the same profile, we
get $U_L \sim D_L$, where
$( U, D ) _{ L }$ denote unitary transformations to go from interaction to mass basis
for LH up and down-type quarks, respectively.
Using $U_L^{ \dagger}D_L = V_{ CKM}$ then gives $\left( U_L \right)_{ 23 } \sim V_{ ts }$
and $\left( U_L \right)_{ 13 } \sim V_{ td }$. 
Combining the above information on left-handed (LH) mixing angles and profile of
$(t,b)_L$, $t_R$ with
the observed quark masses, we can estimate the size of profiles of all the quarks
near the TeV brane
and hence the right-handed (RH) mixing
angles as well
(see reference~\cite{aps} for details).
We find 
$\left( U_R \right)_{ 23} \sim 0.1$ and
$\left( U_R \right)_{ 13} \sim 10^{-3}$,
where
$U_{R}$ denote unitary transformations for RH up-type quarks.

Thus we find:
\begin{eqnarray}
{\cal L}^t_{\rm FC }
& \ni & \Big( g_1 \bar{ t_R }  \gamma_{ \mu }
c_R +
g_2 \bar{ t_L } \gamma_{ \mu } c_L  
\Big) Z^{ \mu } g_Z \,,
\label{tcZ}
\end{eqnarray}
with 
%
%
\begin{eqnarray}
g_{ 1,  2}  \sim    
\Big[
5 \cdot 10^{ -3 } \frac{ \left( U_R \right)_{ 23 } }{ 0.1 },
4 \cdot 10^{ -4 } \frac{ \left( U_L \right)_{ 23 } }{ 0.04 } 
\Big]
\left( \frac{ 3 \; \hbox{TeV} }{ m_{ KK } } \right)^2, 
\end{eqnarray}
and similarly for $\bar{t}uZ$ couplings which are further suppressed. 
Note that 
the 
above
models 
%
%
makes
%
%
a sharp prediction that 
top flavor-violation is mostly right handed.

Next, we consider radiative processes which require chirality flip
and hence result from loop diagrams.
The dominant contributions involve
Higgs and KK fermion in the loop, since the KK fermions
have larger couplings to Higgs than the SM ones:
\begin{eqnarray}
{\cal L}^t_{ FC}
& \ni &\frac{ m_t }{ m_{ W }^2 }\,\left({ \sqrt{ 4 \pi \alpha_{em} },\, g_s }\right) 
\left(F^{ \mu \nu },\,G^{ \mu \nu }\right)\times \nonumber \\ && \bar{t} \sigma_{ \mu \nu }  
  \left( 
C^t_{ 7\gamma,\,8G } P_L + 
C^{\prime\,t }_{ 7\gamma,\,8G } P_R 
\right)
c  \nonumber\,, \\
\end{eqnarray}
where $ F^{ \mu \nu }(G^{ \mu \nu} )$ is the photon (gluon) field strength.
Thus we find
\begin{eqnarray}
C^{ \prime\, t}_{ 7 \gamma,  8G } & \sim & 
\frac{ m_W^2 }{ m_{ KK }^2 }\, \frac{ \lambda_{ 5D } ^2 }{ 16 \pi^2 } 
\left( U_R \right)_{ 23 }\,.
\end{eqnarray}
For the operator with $t_R$, $C^t_{ 7\gamma, 8G }$, 
replace $\left( U_R \right)_{ 23 }$ by $\left( U_L \right)_{ 23 }$
which is further suppressed.

\mysection{Experimental Signals: LHC}
At the LHC $\sim 10^8$ top quark pairs will be produced,
which will allow to search for FCNC top decays with a significantly improved
sensitivity~\cite{atlas}. 
The $tcZ$ coupling in Eq. (\ref{tcZ})
results in 
\begin{eqnarray}
\hbox{ BR} \left( t \rightarrow c Z \right) & \sim &
10^{ -5}
\left( 
\frac{ 3 \; \hbox{ TeV } }{ m_{ KK } } \right)^4
\left( \frac{ \left( U_R \right)_{ 23 } }{ 0.1 } \right)^2.
\end{eqnarray}
Here and below the quantities in parentheses are
$O(1)$ for natural regions of parameter space.
With $100$ fb$^{-1}$ luminosity, the expected upper limit
on BR $\left( t \rightarrow c Z \right)$ is 
$\sim$ a few $10^{ -5}$~\cite{atlas}.
Thus, we see that the (relatively) huge
BR$( t \rightarrow c Z)$ in this model,
much larger than the expectation from the SM
of $\approx 10^{-13}$\cite{ehs91}, is on the edge of current LHC sensitivity,
providing 
a motivation to refine the analysis since
an improvement by an order of magnitude will definitively test this framework.
Also, with enough statistics, angular
analysis will be able to distinguish between LH or RH coupling in $tcZ$~\cite{topPolar}:
the above models predicts that RH coupling dominates.
At the LHC, 
$q \bar{q} \rightarrow tc$ (single top production) 
via $tcZ$ coupling or direct KK $Z$ exchange is 
likely to be overwhelmed by the large background~\cite{Tait}.
However, similar to KK $Z$, there are also flavor violating couplings to 
the KK {\em gluon} which can give observable effects 
in $q \bar{q} \rightarrow tc$
via KK gluon exchange (see reference~\cite{burdman2}).

The dipole
operators give
\begin{eqnarray}\hspace*{-.354cm}
\hbox{ BR}\left( t \rightarrow c \gamma, G \right)
&\sim & 10^{ -10, -9 } \times \nonumber \\
&&   \hspace*{-.254cm}  \left( \frac{ 3 \hbox{ TeV } }{ m_{ KK } }
\right)^2 
\left( \frac{ \left( U_R \right)_{ 23 } }{ 0.1 } \right) 
\left( \frac{ \lambda_{ 5D } }{4} \right)^4,
\end{eqnarray}
dominated by LH operator.
Thus, again we see that BR$(t \rightarrow c \gamma, G)$
in this model is much larger than in the SM\cite{ehs91},
but still too small to be observed: the sensitivities 
at the LHC are
BR$\left( t \rightarrow c \gamma, G \right) \sim 10^{ -5, -4 }$~\cite{atlas}.

{\bf ILC}. 
Indeed the FC-$Ztc$ effective interaction, Eq.(3) has the capacity to
also lead to a striking and clean signature via the
reaction: $e^+ e^- \to t \bar c$ accessible to the  
ILC.
One finds that 
\begin{eqnarray}
  R_{tc}=
  \frac{\zeta_{tc} (a^2_{Ztc} + b^2_{Ztc})(a^2_{Zee} + b^2_{Zee})}
  {\left[(1-m_Z^2/s) 4 \pi \alpha_{ em } \right]^2}\,,
\end{eqnarray}
where
$R_{tc} = \frac{\sigma (e^+ e^- \to [t \bar c + c \bar t])}{\sigma
(e^+ e^- \to \gamma \to \mu^+ \mu^-)}$, 
$\zeta_{tc}=\frac{9}{2}y_c^2 y_t [1 +\frac{y_c}{3y_t}]$,
$y_{c,t} =$   [energy of the charm,top quark/energy of the $e^-$ or
$e^+$] and
$a$'s, $b$'s 
are the coefficient of vector
and axial pieces respectively [$a_{ Ztc }, \; b_{ Ztc } = g_Z(g_1 \pm g_2) / 2$].

The above cross-section is from $tcZ$ coupling and is dominant at low
energies. 
Using the couplings given above and
dimensional
analysis, we can show that at
higher energies, namely, $\sqrt{s} \gsim 
m_Z \,\xi \sim 500$ GeV,
direct KK $Z$, $\gamma$ exchange is  more important
and has a different energy dependence than the SM $Z$ exchange 
\cite{Davoudiasl:2001uj}.
This transition in the energy dependence of the
cross-section may be probed
experimentally providing a clear signature for our framework.

Numerically $R_{tc}$ starts being around $2 \times 10^{-5}$ at energies
close to threshold, {\it i.e.} $\approx 200$ GeV, reaching about
$2 \times 10^{-4}$
at higher energies. It is worth stressing again~\cite{ars95}
that at the ILC this reaction leads to very interesting
and unique signal at relatively low energy, {\it i.e.} $\lsim 2m_t$.
Note also the kinematics of these class of  events is extremely
constrained which should help in their identification.
At such center of mass energies, due to its huge
mass, the top quark takes up well over half (in-fact most of) the energy,  
signifying that it is a single top event, with the opposite
side being an essentially massless (charm) jet, in particular, it must not
contain a b-quark.

Another interesting aspect of this class of events is that the RS1
framework with a generic effective interaction, Eq. (\ref{tcZ}),
leads to a sizeable forward-backward asymmetry 
due to one helicity (in this case RH) being dominant. For unpolarized beams, 
we find that
\begin{equation}
  A_{FB}(e^+ e^- \to t \bar c) = 
 \frac{2\ \zeta_{FB}\ a_{Ztc}\,b_{Ztc} \,a_{Zee}\,b_{Zee}}{(a^2_{Ztc} + b^2_{Ztc})(a^2_{Zee} + b^2_{Zee})}\,,
\end{equation}
where  $\zeta_{FB}=\frac{1 +
  (y_c/y_t)}{1+[y_c/(3y_t)]}\,.$
$A_{FB}$ is 
around 7\% at low energies and asymptotically reaches about
11\%. Note that the asymmetry should be be larger with 
polarized beams. 
Furthermore, the sign of the forward-backward
asymmetry distinguishes dominance of RH vs. LH $Z$ coupling:
it is positive
for RH dominating as in the case of 
the above models with a 
warped extra dimension.
At energies above $500\,$GeV we expect additional contributions
from the direct KK $Z$, $\gamma$ exchange to modify the
form of the asymmetry.

The consensus of the 
community is that the ILC should be initially usable with
energies in the range of 200 to 500 GeV and subsequently
it should be able to run at around 1 TeV. Also, the hope is that the
integrated
luminosity will be around 500 fb$^{-1}$ after the first few years of
running~\cite{list06,fms}. If these characteristics are fulfilled then one can
anticipate tens of FC-$tc$ events.

We end with the following brief comments:

(i) Another interesting feature of the flavor-changing $tc$ vertex
in RS1 is that the mixing coefficient, $(U_R)_{23}$, is actually
complex and in general we should expect $O(1)$ CP-odd phase~\cite{aps}.
In this context the expected beam polarization (80\% for electrons
and up-to about 60\% for positrons~\cite{list06,fms}) at the ILC would become
a very valuable probe. Since, at these energies, the final
state CP-even phases are likely to be small, $T_N$ (naive
time-reversal)-even observables such as partial rate asymmetry are
likely to be rather small. But the several momenta available
(in the decay products of the $tc$ complex), 
in addition to the beam polarization, should
allow us to write down many $T_N$-odd observables~\cite{ans92,pr01}
which will
not require final state phases and could be amenable to experimental
study. 

(ii) With regard to the CP-odd phases a concern in the RS1
type scenario is that in fact one naturally expects neutron
electric dipole moment (NEDM) of $O\left( 10^{-25} \hbox{e-cm}
\right)$ which exceeds
existing experimental bounds by about $O(10)$; therefore
there is a CP ``problem"~\cite{aps}. 
However, there can be significant differences in the size of the
CP phases since the ones that enter the NEDM are from different sectors
$D_R,U_R, U_L,D_L$ than the ones which are relevant to this paper 
(which mostly
arise from $U_R,U_L$).

(iii) 
ILC can also have sensitivity to modifications
of flavor {\em preserving} couplings
of top to SM gauge bosons: to $Z$~\cite{snowmass,DePree:2006ah} and to photon (anomalous 
magnetic moment/EDM-form factors: see below)
via $e^+ e^- \rightarrow \bar{t} t$ 
(ILC will do better here than LHC). In addition, there is a modification
of top quark coupling to the Higgs (from that in the SM) due to the 
mixing of
zero and KK fermions mentioned earlier~\cite{tcH}.
There are also direct gauge KK exchanges
modifying $\bar{t} t$ cross-sections at the ILC (from KK $Z$, $\gamma$)~\cite{DePree:2006ah} and at the 
LHC (from KK gluon).
Diagrams similar to those giving $t \rightarrow c \gamma, 
\; \hbox{gluon}$, but without flavor violation,
give anomalous magnetic moment
for top quark and also EDM in the presence of $O(1)$ CP violating phases:
\begin{eqnarray}
d_t & \sim & 10^{-19} \left( \frac{ 3 \; \hbox{ TeV } }{ m_{ KK } } \right)^2
\left( \frac{ \lambda_{ 5D } }{4} \right)^2 
\hbox{e-cm}\,.
\end{eqnarray}

Needless to say, the (CP-conserving) magnetic form-factor
is likely to be dominated by the standard 1-loop
QCD contribution but the CP-violating electric form
factor, originating from the CKM-phase  is expected to be severely 
suppressed as it cannot contribute at 1-EW loop order; therefore
the RS1 contribution of 1-loop order estimated above is much 
larger.  Note also that in this scenario, for $q^2=s \ll m_{KK}^2$,
$d_t$ is essentially a constant (to $O \left( q^2/m_{KK}^2 \right)$).  
It is thus extremely interesting that the ILC with the parameters 
mentioned above should be able to study top electric 
dipole moment 
form factors of $O \left( 10^{-19} 
\hbox{e-cm} \right)$~\cite{ans92,bern92,pr01}. 

\mysection{Conclusions} Summarizing, the framework of warped
extra dimension provides a novel and very interesting
resolution to the Planck-weak {\em and} flavor
hierarchy problem of the SM. It tends to generically
single out the top quark with properties significantly different
from the SM. In particular, the flavor-changing $tcZ$ interactions
could lead to spectacular signatures at the LHC as well as at the ILC that
would be very worthwhile to explore.

\mysection{Acknowledgements}
We thank Gustavo Burdman for discussions.  
G.P. and A.S.
are supported by 
DOE under Contract Nos. DE-AC02-05CH11231(LBL) and 
DE-AC02-98CH10886(BNL).


\begin{thebibliography}{99}
\bibitem{tcH} Such effects
can also result in flavor-{\em violating} $tcH$ couplings
with BR$\left( t \rightarrow c H\right)$ of roughly
the same size as for $t \rightarrow c Z$ for the case of a light Higgs.
\end{thebibliography}
\end{document}